\documentclass[reprint,prd,aps,superscriptaddress,altaffilletter,amssymb,showpacs,nofootinbib,twocolumn,showkeys,a4paper]{revtex4}

\usepackage[dvips]{graphicx} 
\usepackage[usenames]{color}
\usepackage[latin1]{inputenc}
\usepackage[english]{babel}
\usepackage{multirow}
\usepackage{verbatim}
 \usepackage{amsmath,amssymb}
\usepackage{rotating}
\usepackage{slashbox}
\usepackage[usenames,dvipsnames]{xcolor}

\usepackage{subfigure}

\newcommand{\be}{\begin{equation}}
\newcommand{\ee}{\end{equation}}
\newcommand{\bea}{\begin{eqnarray}}
\newcommand{\eea}{\end{eqnarray}}

\newcommand{\ben}{\begin{eqnarray}}
\newcommand{\een}{\end{eqnarray}}

\newcommand{\ogw}{\Omega_{\rm {gw}} h^2}
\newcommand{\ngw}{N_{\rm gw}}
\newcommand{\neff}{N_{\mathrm{eff}}}

\begin{document}
\title{Improved limits on short-wavelength gravitational waves from the cosmic microwave background}
\author{Irene Sendra}
\email{irene.sendra@ehu.es}
\affiliation{Fisika Teorikoa, Zientzia eta Teknologia Fakultatea, Euskal Herriko Unibertsitatea UPV/EHU, 644 Posta Kutxatila, 48080 Bilbao, Spain}
\author{Tristan L.~Smith}\email{tlsmith@berkeley.edu}
\affiliation{Berkeley Center for Cosmological Physics, Department
of Physics, University of California, Berkeley, CA, USA 94720}
\affiliation{ Institute for the Physics and Mathematics of the Universe (IPMU), University of Tokyo, Chiba 277-8582, Japan}
%\date{\today}

\begin{abstract}

The cosmic microwave background (CMB) is affected by the total radiation density around the time of decoupling. At that epoch, neutrinos comprised a significant
fraction of the radiative energy, but there could also be a contribution from primordial gravitational waves with frequencies greater than $\sim 10^{-15}$ Hz. 
If this cosmological gravitational wave background (CGWB) were
produced under adiabatic initial conditions, its effects on the CMB 
and matter power spectrum would mimic massless non-interacting neutrinos. However, with homogenous initial conditions, as one might expect from
certain models of inflation, pre big-bang models, phase transitions and other 
scenarios, the effect on the CMB would be distinct. 
We present updated observational bounds for both initial conditions using the 
latest CMB data at small scales from the South Pole Telescope (SPT) in combination with Wilkinson Microwave Anisotropy Probe (WMAP), current measurements of the baryon acoustic oscillations, and the Hubble parameter. 
With the inclusion of the data from SPT the adiabatic bound on the CGWB density is improved by a factor of 1.7 to
$10^6\ogw \lesssim 8.7$ at the 95\% confidence level (C.~L.), with weak evidence in favor of an additional radiation component consistent with previous 
analyses. The constraint can be converted into an upper limit on the 
tension of horizon-sized cosmic strings that could generate this gravitational wave 
component, with $ G\mu \lesssim 2 \times 10^{-7}$ at 95\% C.~L., for string tension 
$G\mu$. The homogeneous bound improves by a factor of 3.5 to $10^6\ogw\lesssim1.0$ at 95\% C.~L., with no evidence for such a component from current data.
\end{abstract}
\maketitle

\section{Introduction}

A primordial gravitational wave background may have been 
generated by processes taking place in the early Universe. Given the extremely small cross section of gravitational waves (GWs), they can probe deep into the early Universe, and provide a unique window to explore its evolution.
The exact mechanisms of production of primordial GWs are still under investigation. There are many theoretical models
that predict them: not only through quantum fluctuations during inflation 
\cite{Birrell1984, Abbott1984, Rubakov1982}, but also from cosmic 
strings 
\cite{Damour2000,Damour2001,Siemens2006, Siemens2007, Olmez2010, Smith2006a}, 
causal mechanisms from phase transitions \cite{Hogan1983, Witten1984, Hogan1986,Grojean2007, Apreda2002}, ekpyrotic models \cite{Khoury2001,Khoury2002} 
or pre big-bang theories \cite{Gasperini1993,Enqvist2002}. 
Such backgrounds will involve GWs with wavelengths that extend 
up to our present cosmological horizon, giving a lowest
 observable frequency limit of $\sim 10^{-17}-10^{-16}$ Hz. 

A measure of the amplitude of the cosmological gravitational-wave background (CGWB) at low frequencies can be obtained from constraining a possible tensor-mode contribution to the large-scale temperature and polarization fluctuations in the cosmic microwave background (CMB) \cite{Fabbri1983,Starobinsky1985,Kamionkowski1997,Seljak1997}. Recent results from the Wilkinson Microwave Anisotropy Probe (WMAP) satellite limit the amplitude of tensor fluctuations, quantified by the tensor-to-scalar ratio, to $r<0.20$  at the 95\% confidence level (C.~L.) \cite{Komatsu2011}, which translates to $\ogw < 10^{-14}$ at frequencies $\sim 10^{-17}-10^{-16}$ Hz. At higher frequencies, larger than $\sim 10^{-10}$ Hz, a bound on 
the CGWB can be obtained through big-bang nucleosynthesis (BBN). Gravitational 
waves at these frequencies would contribute to the total radiative 
energy density at the time of nucleosynthesis, thus a measure of the 
light-element abundances can place a constraint on the CGWB. Current observations set an upper limit on the CGWB energy density at this frequency range of $10^6\ogw \lesssim 8.0$ \cite{Cyburt2005}. At the same time, pulsars act 
as natural gravitational wave detectors \cite{Jenet2006}, strongly 
constraining the 
amplitude of the GW with the narrow range of frequencies, 
$f \sim 10^{-9}-10^{-8}$ Hz, to $10^6\ogw < 10^{-2}$ \cite{RVanHaasteren}. 
Large scale interferometers for gravitational wave detection 
(e.g., LIGO \cite{Abramovici1992}, LISA \cite{Shaddock2009}, 
VIRGO \cite{Acernese2002}) are also looking for gravitational wave signals. A recent bound of $10^6\ogw < 6.9$ 
has been obtained at $10^2$ Hz from the Laser Interferometer Gravitational
Wave Observatory (LIGO) \cite{Abbott2009}.

In this work we estimate constraints on the CGWB at frequencies larger than $\sim 10^{-15}$~Hz through their effect on the angular power spectrum of the CMB. This 
component behaves as a non-interacting relativistic fluid, and so would modify the CMB power spectrum in the similar way as 
adding extra neutrino species \cite{Smith2006}. Given the $\sim 2\sigma$ hint of a neutrino number excess seen by both the South Pole Telescope (SPT) and Atacama Cosmology 
Telescope (ACT) experiments \cite{Keisler2011,Dunkley2011}, it is timely to consider alternatives beyond the standard three neutrino 
species. Constraints on this GW background 
were first presented in \cite{Smith2006}, and updated with WMAP 7-year data in \cite{Komatsu2011}. We now make use of the latest CMB data from WMAP and SPT, which 
currently give the tightest constraints on the number of effective neutrinos \cite{Keisler2011}. We then use the improved limits to bound the tension of cosmic 
strings that could generate this gravitational wave background. As in \cite{Smith2006}, we do not restrict our study to the case where the GW where produced under 
adiabatic initial conditions, but also study homogeneous initial conditions. In Section \ref{tb} we introduce the theoretical background, in Section \ref{od} we 
present the observational data used, and finally, in Sections \ref{res} and 
\ref{con} we show the results obtained and discuss our conclusions.

\section{Theoretical background}\label{tb}

The effective number of neutrino species, $N_{\mathrm{eff}}$, represents the energy density stored in relativistic components (radiation) as
\be
\rho_{\rm rad}=\rho_\gamma+\rho_\nu+\rho_x=\left[1+\frac{7}{8}\left(\frac{4}{11}\right)^{4/3}N_{\mathrm{eff}}\right]\rho_\gamma,
\ee
with $\rho_\gamma$, $\rho_\nu$, and $\rho_x$ the energy densities of photons, 
neutrinos, and possible extra radiation components, respectively. In this expression, all non-photon energy density is expressed 
in terms of an effective number of neutrino species, $\neff$. Any additional radiation density, $\rho_x$, would then typically correspond to a non-integer value of $\neff$.
The largest effect of increasing the radiation density on the CMB comes from a decrease in  
redshift of matter-radiation equality, $z_{eq}$, which follows from the relation 
\be
N_{\mathrm{eff}}=3.04+7.44\left(\frac{\Omega_mh^2}{0.1308}\frac{3139}{1+z_{eq}}-1\right),
\ee
for matter density $\Omega_mh^2$ \cite{Komatsu2009}. An increase in $N_{\mathrm{eff}}$ also affects the acoustic oscillations in the 
primordial photon-baryon plasma, leading to additional damping and a phase shift in the acoustic peak positions of the 
CMB (see \cite{Bashinsky2004}). As noted in \cite{Komatsu2011}, the matter density can be better constrained by combining observations of 
the CMB with late-time distance measurements from baryon acoustic oscillation (BAO) data, and a measurement of the Hubble parameter, $H_0$. 
These allow the equality redshift to be better measured, improving constraints from small-scale CMB data on $N_{\mathrm{eff}}$. 

If we assume that there are three standard neutrino flavors, $N_{\mathrm{eff}}=3.046$ (with the small correction due to finite temperature QED effects and neutrino flavor mixing \cite{Mangano2005}), 
and any measured excess would imply an extra relativistic component. Current small-scale CMB data 
show a slight preference for $\neff>3$ at 95\% C.~L., with $\neff=4.56\pm1.5$ measured from the Atacama Cosmology Telescope and $\neff=3.86\pm0.84$ with South Pole Telescope data \cite{Keisler2011} in combination 
with WMAP data, measurements of the BAO, and $H_0$ \cite{Dunkley2011}.  Combining both 
ACT and SPT (along with WMAP and other probes of large-scale structure) increases the significance to $\neff = 4.0 \pm 0.58$ \cite{Smith:2011es}.
As proposed in \cite{Smith2006}, such an excess could be interpreted as a contribution of 
gravitational waves to the radiation density, instead of neutrinos; the effects on the CMB and the matter power spectra are similar
in both cases. Since gravitational waves with wavelengths shorter than the sound horizon at decoupling behave as free-streaming massless particles, a limit on additional relativistic radiation can be translated into an upper limit on the energy density of the CGWB 
for frequencies larger than $\sim 10^{-15}$ Hz.  One can relate the effective `number of 
gravitational-wave' degrees of freedom, $\ngw \equiv N_{\mathrm{eff}}-3.046$, to the CGWB energy density \cite{Maggiore2000}:
\be
\ogw\equiv \int_{0}^{^{\infty}} d\left( \ln{f}\right) h^2 \Omega_{gw}(f) = 5.6\times10^{-6}\ngw.
\label{eq:int_Ogw}
\ee

The different processes which may produce a CGWB lead to different initial conditions for perturbations in the CGWB fluid. 
Adiabatic initial conditions may arise from the 
incoherent superposition of cusp bursts from a network of cosmic strings or superstrings (see \cite{Siemens2007,Olmez2010} for further details). Bounds on the energy 
density of this radiation can then be translated into an upper limit on the string tension, $G\mu$, for a given model 
\cite{Vilenkin1981,Damour2000,Damour2001,Siemens2006, Siemens2007, Olmez2010}.  Following the procedure given in \cite{Olmez2010} for the analytical approximation of the CGWB produced by strings, we can relate our limit on the GW energy density to a bound on $G\mu$ and the string reconnection probability, $p$, which is 1 for cosmic strings and $<1$ for superstrings.  String loops which are smaller 
than the causal horizon produce a CGWB with 
\begin{equation}
\Omega_{\rm gw}(f) \approx 5 \times 10^{-2} G \mu / p
\label{eq:short_spectrum}
\end{equation}
whereas the CGWB produced by horizon sized loops scale as 
\begin{equation}
\Omega_{\rm gw}(f) \approx 3.2\times10^{-4} \sqrt{G \mu} / p.  
\label{eq:long_spectrum}
\end{equation}

An alternative scenario, noted in \cite{Smith2006}, may arise when the CGWB is produced by quantum fluctuations during a period of inflation.  In this case, the perturbations may be non-adiabatic, differing from those of other species produced though the decay of the inflation.
We  
re-examine the dependence on initial conditions by also considering `homogeneous' initial conditions, as in \cite{Smith2006}. These have no initial primordial 
perturbations in the CGWB energy density in the Newtonian gauge, thus the curvature perturbation disappears when the CGWB energy density dominates, and, in addition, 
it approaches the adiabatic case when the CGWB energy density vanishes. 
 
\section{Data and methodology}\label{od}

\begin{figure}[t!]
\includegraphics[width=0.45\textwidth]{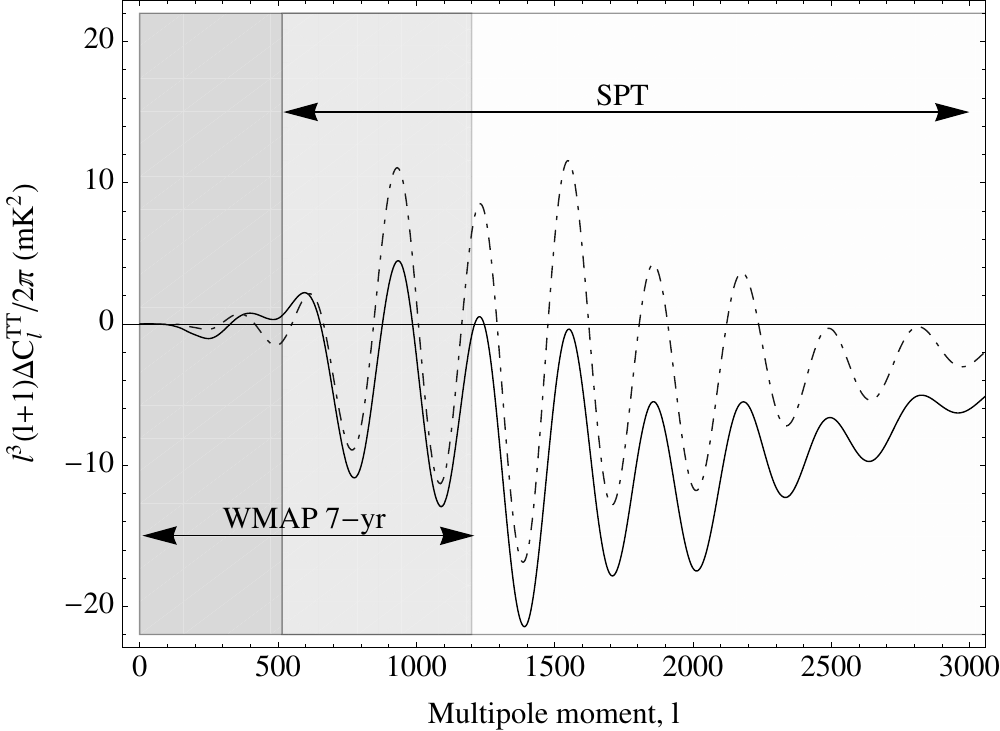}
\caption{The difference between the temperature power spectrum, $\Delta C_{l}^{TT}$, between a model with $\ngw = 0$ and $\ngw = 1$ with the other cosmological parameters chosen so that the two models fit the WMAP-7 data equally well.
The thick-line model has adiabatic initial conditions, the dot-dashed homogeneous initial conditions. The angular 
range of the SPT and WMAP $7$-year CMB data are shown.}
\label{clscomp}
\end{figure}

We use the latest measurements of the CMB power spectrum from the WMAP 7-year data release \cite{Larson2011} and SPT \cite{Keisler2011}, which extend the previous set of data to higher multipole moments ($\ell_{\rm max}\sim3000$). These data considerably improve constraints on $N_{\mathrm{eff}}$, which will be translated into tighter bounds on $\ogw$.  To illustrate the sensitivity of these new data to changes in $\neff$, in Fig.~\ref{clscomp} we show the difference between a model with $\ngw = 0$ and $\ngw = 1$ with the other cosmological parameters chosen so that the two models fit the WMAP-7 data equally well.  It is clear that the data from SPT covers the range where the power spectrum is sensitive to a variation of $\ngw$.

We estimate the bounds on $\ngw$ by varying the number of effective gravitational-wave degrees of freedom, $\ngw$, imposing a prior that $\ngw>0$ and assuming the standard three non-interacting massless neutrino species and by
marginalizing over the standard six $\Lambda$CDM parameters: the baryon energy density in units of the critical energy density, $\Omega_{\rm b} h^2$, the energy density in cold dark matter in units of the critical energy density, $\Omega_{\rm DM} h^2$, the angular size of the first acoustic peak $\theta$, the optical depth to the surface of last scattering, $\tau$, the scalar spectral-index, $n_s$, and the amplitude of the scalar power-spectrum, $A_s$. To include the SPT data we follow the prescription presented in \cite{Keisler2011}, marginalizing over three additional parameters describing power from foregrounds and the Sunyaev-Zel'dovich effect. We use a Markov Chain Monte Carlo (MCMC) method to estimate the probability distribution, using the CosmoMC\footnote{http://cosmologist.info/cosmomc/} software package, see \cite{Lewis2002,cosmomc_notes} with a modified version of CAMB\footnote{http://camb.info/} \cite{Lewis2000} which includes the homogeneous initial conditions for CGWB perturbations in the conformal Newtonian gauge. We combine these CMB measurements with distance measurements from Baryon Acoustic Oscillation (BAO) data measured at $z=0.2$ and $z=0.35$ from SDSS and 2dFGRS presented in \cite{percival2007}. We also include a prior of $H_0=74.2\pm 3.6$ km/(s Mpc) on the Hubble constant \cite{Riess2009}. The CMB temperature and polarization spectra are obtained with CAMB for each cosmology, and then passed to CosmoMC to obtain the joint likelihood of WMAP, SPT, BAO, and $H_0$, as described in \cite{Keisler2011}\footnote{Likelihood code is available at http://lambda.gsfc.nasa.gov/}.

\section{Results}\label{res}

\begin{figure}
\includegraphics[width=0.5\textwidth]{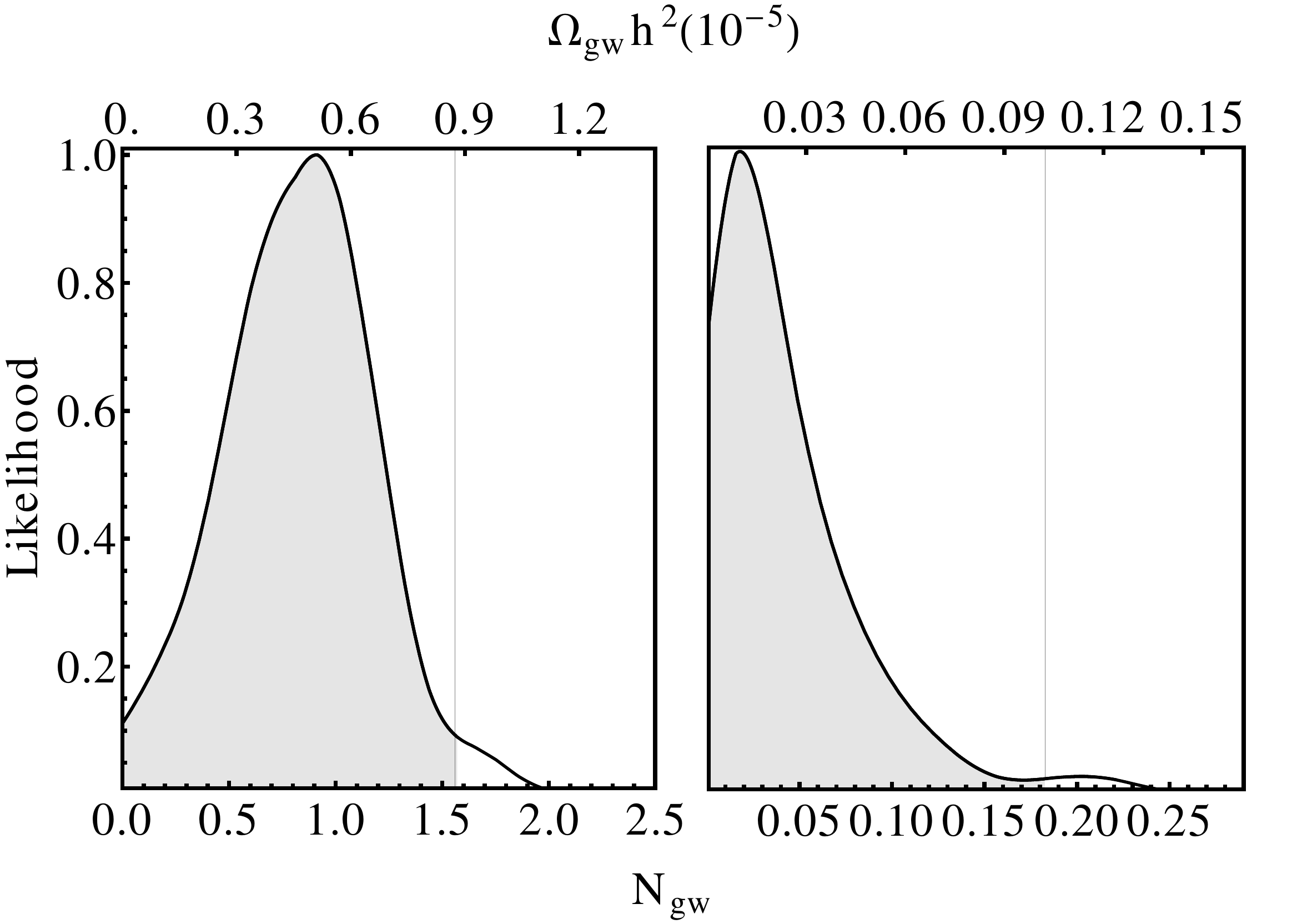}
\caption{\label{fig:comp} Marginalized likelihoods for $\ngw$ and CGWB energy density, $\ogw (\times 10^{-5})$, 
for adiabatic (left) and homogeneous (right) initial conditions, where the horizontal line and the colored/shaded region indicate the 95 \% C.L. limit.}
\end{figure}

As shown in Fig.~\ref{fig:comp} and Table \ref{tbl:results} the new combination of CMB data gives an upper limit of $\ngw<1.56$, or $10^6 \ogw < 8.71$,  at $95 \%$ C.~L. for adiabatic initial 
conditions which improves the previous upper bounds on $\ogw$ given 
in \cite{Komatsu2011} by a factor of 1.7.
A non-zero relativistic excess is weakly favoured at 95\% C.~L., with the best-fit value $\ngw=0.82\pm0.36$. Homogeneous initial 
conditions for GWs are now more strongly excluded, with $\ngw<0.18$, $10^6\ogw < 1.02$, an improvement on the previous constraint in \cite{Smith2006a} by a factor of 3.5. In this case the degeneracy between the effects of CGWB and 
neutrinos is broken, giving us the stronger constraint. These bounds are now competitive with those given by BBN, 
$10^6\ogw < 8.0$ \cite{Cyburt2005} for $f \gtrsim 10^{-10}$~Hz and LIGO,
$10^6\ogw < 6.9$  at $f \approx 10^2$~Hz \cite{Abbott2009}, but with the benefit that 
this upper bound can be extended
to frequencies as low as $10^{-15}$~Hz. 

\begin{figure}
\includegraphics[width=0.45\textwidth]{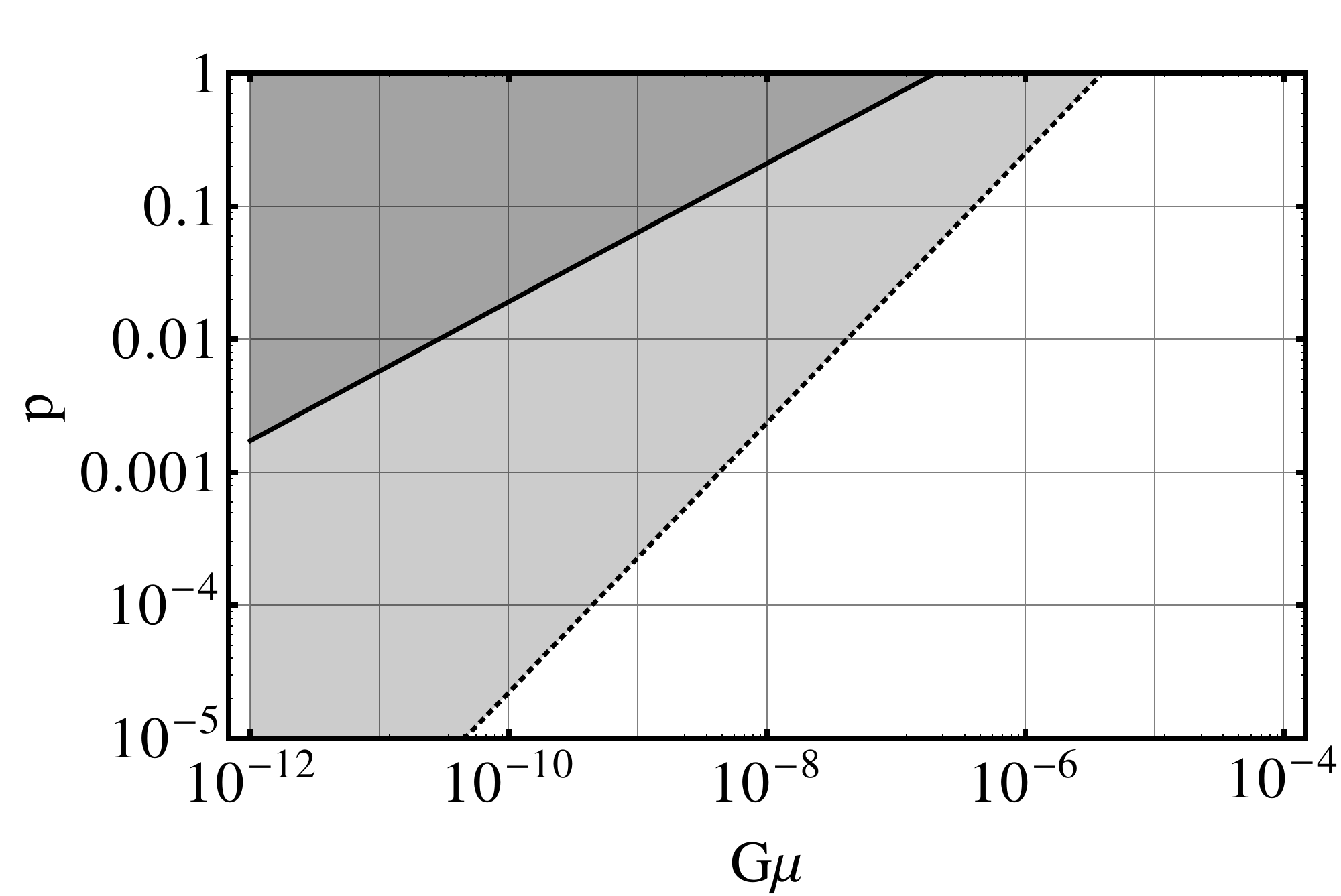}
\caption{\label{fig:gmup}  The shaded region shows the 95\% C.~L. region which is allowed by constraints on the CGWB from the CMB in the $p-G\mu$ plane for horizon-sized (solid), and subhorizon-sized (dashed) cosmic string and superstring models with string tension $G\mu$ and reconnection probability $p$, given a GW 
energy density level $\ogw=8.7\times 10^{-6}$. For $p=1$, the CMB power spectrum limits $G\mu \lesssim 10^{-6}-10^{-7}$.}
\end{figure}

Our constraint to the integrated CGWB energy density with adiabatic initial conditions can be translated into a constraint on 
cosmic string networks, as shown in Fig.~\ref{fig:gmup}.  In order to do this we use Eq.~(\ref{eq:int_Ogw}) and integrate the spectra, given in Eqs.~(\ref{eq:short_spectrum}) and (\ref{eq:long_spectrum}).  The lower-bound to the CGWB spectrum produced by the string network is 
$f_{\rm min} \sim 3.6 \times 10^{-18}/(G\mu)$ Hz for horizon-sized string loops and $f_{\rm min} \sim \alpha^{-1} z_{\rm eq}^{1/2} H_0$ for 
sub-horizon sized string loops where $z_{\rm eq} \simeq 3400$ is the redshift at matter-radiation equality \cite{Olmez2010}.  The upper-bound is given by the horizon size at the 
time of the phase-transition which produced the network, $f_{\rm max} \sim \alpha^{-1}H(T_{\rm pt}) = \alpha^{-1}T_{\rm pt}^2/M_{\rm pl} = \alpha^{-1}M_{\rm pl} G\mu$, where $T_{\rm pt} = \sqrt{G\mu} M_{\rm pl}$ is the 
temperature of the phase transition,  $M_{\rm pl}=1/\sqrt{8\pi G}$ is the Planck mass, and $\alpha = 1$ for horizon-sized string loops  \cite{Olmez2010}.
With this we find that 
for subhorizon-sized loops our constraint becomes, 
\begin{equation}
\frac{G\mu}{p} \ln\left(\frac{ G\mu M_{\rm pl} }{H_0z_{\rm eq}^{1/2}}\right) \lesssim 3 \times 10^{-4},
\end{equation}
and for horizon-sized loops we have, 
\begin{equation}
\frac{\sqrt{G\mu}}{p} \ln\left(\frac{(G\mu)^2 M_{\rm pl}}{3.6\times10^{-18}\ {\rm Hz}}\right) \lesssim 5 \times 10^{-2}.
\end{equation}
For $p=1$ we obtain $G \mu \lesssim 2 \times 10^{-7}$ for horizon-sized string loops and $G \mu \lesssim 2.5\times 10^{-6}$ for subhorizon loops.  String tensions 
this large have already been excluded  through directly limiting a string contribution to the CMB power spectrum \cite{Dunkley2011, Bevis2008, Battye2010,Urrestilla2011,Dvorkin2011}, $G\mu \lesssim 10^{-6}-10^{-7}$. However, for lower reconnection probability, e.g.\ $p=0.1$, the $\ngw$ bound tightens to $G \mu \lesssim 2.4 \times 10^{-9}$ for horizon-sized string loops. Further study is required to investigate constraints considering both the string contribution
to the CMB power spectra, and additional relativistic degrees of freedom.
 
\begin{table}[!tb]
\label{tabresult}
\caption{Upper limits on $\ngw$ and $\ogw$ at $95\%$ C.L for the adiabatic and homogeneous primordial initial conditions. 
\label{tbl:results}}
\centering
\begin{tabular}{c|cc}
\hline \hline
 & $\ngw$ & $10^6\ogw$ \\\hline 
Adiabatic 95\% upper limits & $< 1.56$ & $< 8.71$      \\ 
Adiabatic 68\% range & $0.82\pm0.36$ & $4.8\pm0.2$    \\ 
Homogeneous 95\% upper limits &$< 0.18$ &  $< 1.02$  \\ \hline
\end{tabular}
\end{table}

\section{Conclusions}\label{con}

We have used recent CMB data at small angular scales from SPT, in combination with data from WMAP-7 year and the latest measures of $H_0$ and BAO, to 
constrain a possible cosmological gravitational wave background with frequencies greater than $10^{-15}$ Hz. Recent measurements of the 
Silk damping tail from SPT improves limits on the CGWB density by about a factor of two.  We note that the inclusion of additional measurements of the small-scale CMB, such as from the Atacama Cosmology Telescope, do not significantly improve the upper-limits presented here \cite{Smith:2011es}.  

In agreement with other analyses of similar data we find that a non-zero relativistic energy density with adiabatic initial 
conditions is preferred at the 95\% level \cite{Dunkley2011,Keisler2011,Smith:2011es}.  This additional energy density 
could be attributed to gravitational waves with frequencies greater than $\sim 10^{15}$~Hz, or some other relativistic species (see, e.g., \cite{Smith:2011es}). If we attribute this extra relativistic energy density to gravitational waves produced by a 
cosmic string network it is interesting to note that for $p=1$ the upper limits to the CGWB presented here are competitive with direct CMB constraints on the the string tension, both now requiring $G\mu \lesssim 10^{-6}-10^{-7}$.

We have also explored how constraints on the CGWB depend on the initial conditions.  Besides the standard adiabatic initial 
conditions, we placed constraints 
on the CGWB energy with `homogenous' initial conditions \cite{Smith2006a}. The additional information on the small-scale CMB from SPT improves upon the previous upper-limit by 
more than a factor of 3.5. As opposed to an additional adiabatic relativistic component, the data does not prefer any additional homogeneous relativistic energy density. 

 Finally, the situation will be greatly improved with near-future CMB observations. For example, the Planck satellite will increase the sensitivity of these constraints by a factor of 5 for adiabatic initial conditions and a factor of 2 for homogeneous initial conditions \cite{Smith2006a}. 

\begin{acknowledgements}
IS is grateful to Joanna Dunkley for proposing this problem and her devoted guidance and hospitality during the author's stay 
at the University of Oxford.  IS also acknowledges Jon Urrestilla for his help with cosmic strings and other fruitful discussions
and Ruth Lazkoz for her limitless support and good advice.  We thank Xavier Siemens for useful discussions about the gravitational wave backgrounds produced by cosmic string and superstring models. 
IS holds a PhD FPI fellowship contract from the
Spanish Ministry of Economy and Competitiveness. She is also supported by the mentioned ministry
through research projects FIS2010-15492 and Consolider
EPI CSD2010-00064,  by the Basque Government through research project
GIU06/37 and
special research action KATEA, and by
the University of the Basque Country UPV/EHU under
program UFI 11/55 and research project Etorkosmo.  TLS was supported by the Berkeley Center for Cosmological Physics and the Institute for the Physics and Mathematics of the Universe (IPMU) at the University of Tokyo. 
\end{acknowledgements}

\bibliography{tesis}

\end{document}